\documentstyle[aps,eqsecnum,epsf,prd]{revtex}
\begin{document}
\twocolumn[
\preprint{\vbox{
\hbox{UCSD/PTH 94--19}
\hbox{hep-ph/9506273}
}}
\title{Flavor Symmetry Breaking in the $1/N_c$ Expansion}
\author{Jin Dai, Roger Dashen, Elizabeth Jenkins and Aneesh V.~Manohar}
\address{Department of Physics, University of California at San Diego,\\
La Jolla, CA 92093}
\date{June 1995}
\maketitle
\widetext
\vskip-1.5in
\rightline{\vbox{
\hbox{UCSD/PTH 94--19}
\hbox{hep-ph/9506273}
}}
\vskip1.5in
\begin{abstract}
The breaking of flavor \thinspace $SU(3)$ symmetry in the axial couplings
and magnetic moments of baryons is analyzed in the $1/N_{c\text{ }}$%
expansion. A simple meson loop graph which is known to be of order $\sqrt{m_s}$
and leading order in $1/N_c$ correctly predicts the pattern of symmetry
breaking in the magnetic moments. It is, however, not possible to use this
graph to predict the magnitude of the breaking. The situation with the axial
couplings is less clear. In this case the breakings are relatively small and do
not appear to follow an obvious pattern. Nevertheless there is a clear
indication that, with symmetry breaking taken into account, the $F/D$ ratio
(defined in the presence of $SU(3)$ breaking)
is considerably less than the $SU(6)$ value of $2/3$.  With sizeable
uncertainty, we find $F/D\approx 0.44$.  The quantity $3F-D$
which is relevant for the analysis of spin-dependent deep inelastic
scattering is considerably smaller than the $SU(6)$
value of unity.  The new value, $3F-D=0.27\pm0.09$, is consistent with
vanishing strange-quark spin in the nucleon.
\end{abstract}

]

\narrowtext

\section{INTRODUCTION}

The $1/N_c$ expansion has proved to be quite useful in understanding the
spin-flavor structure of baryons in QCD \cite{djm2,dm,EJ,JM,JL}.
In the
flavor symmetry limit, a structure similar to $SU(6)$ emerges \cite
{djm2,dm,EJ,JM,JL,G-S,LMR,CGO,CGO2}. This is not, however, a statement that
there is an $SU(6)$ symmetry in the Lagrangian or that the quarks are
nonrelativistic. Rather, it is the simplified dynamics \cite{GH,EW} of QCD
at large $N_c$ that produces patterns similar, but not exactly equal to, the
predictions of $SU(6).$ In this paper we focus on flavor $SU(3)$ breaking in
the baryon magnetic moments and axial couplings in the $1/N_c$ expansion.
The mathematical groundwork, which involves a considerable amount of group
theory, was set up in a previous paper \cite{djm2} --- here we will
concentrate on the data and on the physical interpretation of results.

The pattern of symmetry breaking displayed by the baryon magnetic
moments is in quite good agreement with that predicted by the simple meson
loop diagram shown
in Fig.~\ref{fig:loop} \cite{JLMS,LMW}. In the chiral limit, this diagram is of
order
$\sqrt{m_s}\,N_{c\text{ }}$, i.e. it is of order $\sqrt{m_s}$ rather than the
naively
expected $m_s$ and is leading order in $N_{c\text{ .}}$
However, a naive evaluation of the diagram in
Fig.~\ref{fig:loop} predicts symmetry breaking which is, in magnitude,
about a factor of
two too large. As pointed out in Ref. \cite{JLMS}, this is almost
certainly
due to the fact that the diagram involves meson momenta on the order of $M_{K%
\text{ }}$which are too large for the validity of chiral perturbation theory
and, consequently, the numerical value of the diagram cannot be taken
seriously.  Nevertheless, the group theoretic structure suggested by this
diagram is in rather impressive agreement with the data.

For the axial couplings the situation is less clear. Here the data are the
widths of the decuplet baryons, converted to axial couplings through the
Goldberger-Treiman relation, and the values of $g_A$ extracted from
$\beta$-decays of the octet baryons.
In this case the $SU(3)$ breaking is actually
rather small -- the leading term is order $m_s\,N_c$ rather than $\sqrt{m_s}
\,N_c $ -- and the experimental errors are larger. While we do find what
appears to be a stable fit with an interesting physical interpretation, we
have some concerns about the results. Although we feel that we are not in a
position to make precise and definitive statements about symmetry breaking
in the axial couplings, some trends do emerge. The first is that the pattern
of symmetry breaking is broadly consistent with the $1/N_c$ expansion and the
fact that chiral perturbation theory does not predict a dominant pattern.
The second is the value of the $F/D$ ratio, a quantity which in the presence
of $SU(3)$ breaking must be carefully defined. We parameterize the couplings
in such a way that the matrix elements of the strangeness preserving
$\Delta S=0$
currents (both isovector and isoscalar) between nucleon states are exactly
given by $D$ and $F$, at least through first order in $SU(3)$ breaking.
Using this definition we find $F/D\approx 0.44$ with perhaps a 10\% error.
Although the error is significant, the central value for the $F/D$ ratio
is considerably smaller
than the $SU(6)$ value of $2/3$.  Finally,
the above definition of $D$ and $F$ is used to obtain a new value
for the quantity $3F-D$ which appears in the analysis of
spin-dependent deep inelastic scattering~\cite{EMC,SMC,E142,E143}.
Our new value, $3F-D\approx 0.27\pm0.09$, is
considerably smaller than the $SU(6)$ value of unity or the value of
$\approx 0.6$ obtained from a $SU(3)$ symmetric fit to the data.
The value
of the strange-quark spin in the nucleon extracted from experimental data is
significantly reduced using the new value of $3F-D$ and
is consistent with zero.
The uncertainty in the value of $3F-D$ limits the accuracy of the extraction
of the nucleon's strange-quark spin.
Since one is in a region of parameter space where there is a
large cancellation between $3F$ and $D$, neither this analysis nor any
other is likely to give a highly accurate value for $3F-D$.

We had originally hoped that studying the
magnetic moments would help in the interpretation of the axial couplings ---
in the naive non-relativistic quark model, magnetic moments and axial
couplings tend to behave in the same way. However, as mentioned above, the
actual symmetry breaking in the magnetic moments appears to come largely
from a single diagram which has no analog in the axial couplings. The
reason we can make some progress in the analysis of the axial couplings
is that the $1/N_c$ expansion predicts
that some $SU(3)$ breaking operators are suppressed, and it relates the
octet and decuplet axial couplings so that one has more experimental
input to constrain the fit.

The baryon axial currents in the presence of $SU(3)$ breaking also have
been studied recently by Ehrnsperger and Sch\"afer~\cite{ES} who assume
that the $SU(3)$ breaking is proportional to the baryon masses, and by
Lichtenstadt and Lipkin~\cite{LL} using a model. Both calculations give
results which are similar to those obtained here: $SU(3)$ breaking
lowers the value of $F/D$ and of $3F-D$.

The paper is organized as follows. In the next section the operator analysis
of Ref. \cite{djm2} is summarized, with some additional mathematical details
given in an appendix. The operator analysis leads to a seven
parameter fit to the hyperon $\beta$-decays and decuplet pionic decays,
with coefficients listed in Table~\ref{AxiC}, and a seven parameter fit to the
baryon magnetic moments with coefficients listed in Table~\ref{MagC}.
The reader not interested in the technical details of the $1/N_c$
expansion can skip directly to Sections~\ref{sec:skip} and~\ref{sec:mag},
which present the
analysis of the axial couplings and the magnetic moments, respectively.

\section{Operator Analysis}

Confining ourselves to the physically interesting case of three light
flavors, the lowest-lying baryon states fall into a representation of the
spin-flavor group $SU(6)$. At the physical value $N_c=3$, this is the
familiar {\bf 56 }dimensional representation of $SU(6)$ while for larger $%
N_c$ the representation becomes more complex, containing spins greater than
$3/2$ and $SU(3)$ representations bigger than the {\bf 8} and {\bf 10}. The
complexity of these large representations of $SU(6)$ makes a
straightforward application of group theory very tedious at large $N_c$. It was
pointed out in Refs. \cite{djm2,LMR,CGO} that it is easier to focus on the
operators than the states. The magnetic moments, for example, have an expansion
in
operators whose coefficients are inverse powers of $N_c$. Working to a given
order in $1/N_c$, one can truncate the expansion and connect to physics by
evaluating the matrix elements at $N_c=3$. For any representation of $SU(6)$%
, polynomials in the generators $J^i,T^a$ and $G^{ia}$ form a complete set of
operators. For the particular representations relevant to the lowest-lying
baryons, it suffices to keep polynomials through order $N_c$ and, in
addition, there are a number of identities among the polynomials of order
less than or equal to $N_c$. The problem of finding a complete and
independent set of operators for the baryon representations was solved in
Ref. \cite{djm2}.

The results of Ref.\cite{djm2} can be summarized as follows. The basic
building blocks are the operators $J^i$, $T^a$ and $G^{ia}$, where $J^i$ is the
spin, the $T^a$ are the generators of $SU(3)$ and the $G^{ia}$ are the
remaining generators of $SU(6)$. They satisfy the commutation relations
\begin{eqnarray}
\lbrack J^i,J^j] &=&i\,\epsilon^{ijk}J^k, \nonumber\\
\lbrack T^a,T^b] &=&i\,f^{abc}T^c,  \nonumber \\
\lbrack G^{ia},G^{jb}] &=&i\,\delta^{ij}f^{abc}T^c+i\epsilon^{ijk}
\left(\delta^{ab}J^k+d^{abc}G^{kc}\right),  \nonumber\\
\lbrack T^a,G^{ib}] &=&i\,f^{abc}G^{ic},  \nonumber\\
\lbrack J^i,G^{ja}] &=&i\,\epsilon^{ijk}G^{ka},  \nonumber
\end{eqnarray}
and can be defined in terms of the quarks as
\begin{eqnarray*}
J^i &=&q^{\dagger }\frac{\sigma^i}2 q, \\
T^a &=&q^{\dagger }\frac{\lambda^a}2 q, \\
G^{ia} &=&q^{\dagger }\frac{\sigma^i\lambda^a}4 q\text{ .}
\end{eqnarray*}
A complete set of operators can be constructed from polynomials in the
operators $J$, $T$ and $G$.  Because antisymmetric products can be reduced
using the commutation relations, one need only consider products which are
completely symmetric in non-commuting operators. Furthermore, it can be
shown that all products of $T$'s and/or $G$'s in which two flavor indices are
summed over or contracted with $d$ or $f$ symbols can be eliminated in terms
of lower order products.

The way in which large $N_c$ dynamics enters can best be seen through an
example. Let $P^{ia}$ be the operator whose matrix elements between $SU(6)$
symmetric states gives the actual axial couplings of the baryons. It is
spin-one, an octet under $SU(3)$, and odd under time reversal. In the absence
of
$SU(3)$ breaking and neglecting quartic and higher order polynomials, $%
P^{ia} $ can be written as
\begin{eqnarray}
\frac 1 2 P^{ia}&=&a\,G^{ia}+b\,J^iT^a+d\left\{J^2,G^{ia}\right\}\nonumber\\
&&+e\left\{J^i,\left\{J^k,G^{ka}\right\}\right\} \label{symp}
\end{eqnarray}
where $a,b,d$ and $e$ are unknown coefficients. By examining diagrams one
can see that $a$ is order $N_c^0$, $b$ is of order $N_c^{-1}$ and $d$ and $e$
are of order $N_c^{-2}$ . Hence, $P^{ia}$ can also be expressed as
\begin{eqnarray}
\frac 1 2 P^{ia}&=& a^{\prime }\,G^{ia}+b^{\prime }{1 \over N_c}J^iT^a
+d^{\prime }{1 \over N_c^2}\left\{J^2,G^{ia}\right\} \nonumber\\
&&+e^{\prime }{1 \over N_c^2}\left\{J^i,\left\{J^k,G^{ka}\right\}\right\}
\label{su3not}
\end{eqnarray}
where the new coefficients are of order unity for large $N_c$.

If we
consider states whose spin remains fixed as $N_c\rightarrow \infty$ , then
the matrix elements of $J$ never become large. The operators $T$ and $G$ are
more complicated --- for some states in the representation $G$ has matrix
elements of order $N_c$ while $T$ has matrix elements of order unity, but
for other states $G$ has matrix elements of order unity while $T$ has matrix
elements of order $N_c$. There are other states for which both $G$ and $T$
have matrix elements of order $\sqrt{N_c}$. Nevertheless, it is clear that
truncating $P^{ia}$ at order $N_c^{-1}$ is a consistent approximation; the
remaining two terms multiplied by $d$ and $e$ are everywhere smaller than
the first term. Dropping the second term is more problematic -- there are
states for which the matrix elements of $J^iT^a/N_c$ are of the same order
as the matrix elements of $G^{ia}$.  Thus, in general, this term must be
retained.
Finally, note that keeping all four
terms allows for arbitrary values of the four possible $SU(3)$
symmetric couplings of pseudoscalar mesons to the octet and decuplet
baryons. This is an example of the fact that for $N_c=3$ we never have to go
beyond operator products of third order in the generators.

When first order $SU(3)$ breaking is taken into account, $P^{ia}$ contains
pieces transforming according to all $SU(3)$ representations contained in
the product
${\bf 8}\otimes {\bf 8}={\bf 1}\oplus {\bf 8}_A\oplus {\bf 8}_S
\oplus {\bf 10}\oplus \overline{\bf 10}\oplus {\bf 27}$.
We summarize the results of a more detailed analysis presented in
Ref.~\cite{djm2}. The possible spin-one
singlets containing three or fewer generators are
$J^i$ and $J^2J^i$ with the second operator
having a coefficient of order $N_c^{-2}$
relative to that of the first. The only singlet term that we will keep is $%
\delta^{a8}J^i$. The octet operators were discussed in the previous
paragraph. We will drop the $N_c^{-2}$ terms, leaving $d^{ab8}G^{ib}$ and $%
d^{ab8}J^iT^b$ as the octet terms.  (Similar terms with the $d$
symbol replaced by an $f$ symbol are ruled out by time reversal.) We do,
however, have to keep the operator
\begin{equation}
\lbrack J^2,[T^8,G^{ia}]]  \label{offdiag}
\end{equation}
which can be reduced, by use of the commutation relations, to a sum of
operators of the form $\{J^k,G^{jb}\}$ -- this particular sum cannot appear
in the $SU(3)$ limit by time reversal invariance but is allowed in the
case of broken symmetry. This operator receives a coefficient of order $%
N_c^{-1}$ and contributes only to processes which change both spin and
strangeness. The leading operator containing a ${\bf 27}$ is
\begin{equation}
\{G^{ia},T^8\}+\{G^{i8},T^a\}  \label{o27}
\end{equation}
which receives a coefficient of order $N_c^{-1}$. In the next order there are
two operators, $J^i\{T^a,T^8\}$ and $\{J^kG^{ka},G^{i8}\}+\{J^kG^{k8},G^{ia}%
\}$ which have coefficients of order $N_c^{-2}$. The leading operator
containing ${\bf 10}\oplus \overline{\bf 10}\,$ is
\begin{equation}
\{G^{ia},T^8\}-\{G^{i8},T^a\}  \label{o10}
\end{equation}
with a coefficient of order $N_c^{-1}\,$and in the next order one has $%
\{J^kG^{ka},G^{i8}\}-\{J^kG^{k8},G^{ia}\}$ with a coefficient of order $%
N_c^{-2}$. We will keep only the leading ${\bf 27}$ and
${\bf 10}\oplus \overline{\bf 10}$
terms. In Appendix A, it is shown that matrix elements of the higher order
terms are always down by a factor of at least $N_c^{-1}$ relative to matrix
elements of the leading terms.

In terms of $N_s$ and $J_s^i$, the number of strange quarks and the strange
quark spin, defined by
\begin{eqnarray}
G^{i8} &=&\frac 1{2\sqrt{3}}(J^i-3J_s^i)  \nonumber \\
T^8 &=&\frac 1{2\sqrt{3}}\left( N_c-3N_s\right),  \label{defNsJs}
\end{eqnarray}
the leading ${\bf 10}\oplus \overline{\bf 10}$ and ${\bf 27}$ operators are $%
\{G^{ia},N_s/N_c\}$ and $\{T^a,J_s^i/N_c\}$, where we have dropped constants
and terms that simply renormalize the symmetric couplings. A consistent
truncation of $P^{ia}$ valid to first order in $SU(3)$ breaking is therefore
\begin{eqnarray}
\frac 1 2 P^{ia} &=&
\left(a^{\prime }\delta ^{ab}+c_1^{\prime }d^{ab8}\right)G^{ib}
+\left(b^{\prime}\delta ^{ab}+c_2^{\prime }d^{ab8}\right)\frac{J^i T^b}{N_c}
\label{Axi} \nonumber\\
&&+c_3^{\prime }\left\{G^{ia},\frac{N_s}{N_c}\right\}
+c_4^{\prime }\left\{T^a,\frac{J_s^i}{%
N_c}\right\}  \\
&&+c_5^{\prime }\left[J^2,\left[\frac{N_s}{N_c},G^{ia}\right]\right]
+c_6^{\prime }\delta ^{a8}J^i
\nonumber
\end{eqnarray}
where the $c_k^{\prime }$ are of order $SU(3)$ breaking and the scaling with
$N_c$ is explicit. The large $N_c$ expansion yields one further piece of
information: as explained in Ref. \cite{djm2} the coefficients are
constrained by
\begin{equation}
3c_6^{\prime }=c_1^{\prime }+c_2^{\prime }  \label{constrain}
\end{equation}
up to terms of order $N_c^{-1}$. Rearranging terms and absorbing factors of $%
N_c^{-1}$ into the coefficients leads to the form that we will actually use
in fitting data:
\begin{eqnarray}
\frac 1 2 P^{ia} &=&a\,G^{ia}+b\,J^iT^a
+\Delta ^a(c_1G^{ia}+c_2J^iT^a) \nonumber \\
&&+c_3\left\{G^{ia},N_s\right\}+c_4\left\{T^a,J_s^i\right\}  \label{Mag} \\
&&+\frac{\delta ^{a8}}{\sqrt{3}}W^i+d\left\{J^2-\frac 34,G^{ia}\right\}
\nonumber
\end{eqnarray}
where
\begin{eqnarray}
W^i&=&(c_4-2c_1)J_s^i+(c_3-2c_2)N_sJ^i\nonumber\\
&&-3(c_3+c_4)N_sJ_s^i,
\end{eqnarray}
$\Delta _a=1$ for $a=4,5,6$ or $7$ and is zero otherwise, and the unprimed
coefficients are linear combinations of the primed ones. The term involving $%
[J^2,[T^8,G^{ia}]]$ which does not contribute to any observed decay has been
dropped and a term $d\{J^2-3/4,G^{ia}\}$ has been added to allow the $SU(3)$
symmetric parameters\cite{DefCH} $D$, $F$ and ${\cal C}$
to have arbitrary values. Our main interest is to study $SU(3)$ breaking and
adding this extra
symmetrical term keeps symmetry breaking effects from being mixed up with
$1/N_c$ corrections in the symmetric couplings. The
coefficient $d$ is of order $N_c^{-2}$ and is presumably comparable to some
of the other coefficients, e.g. $c_2$ which is of order $\epsilon /N_c$
where $\epsilon \sim 0.3$ is the strength of $SU(3)$ violation.
Note that
the couplings have been parametrized in such a way that only the symmetric
parameters $a$, $b$ and $d$ contribute to processes which take place
entirely in the strangeness zero sector. We define $D$ and $F$ in the presence
of $SU(3)$ breaking by
\begin{eqnarray}
D&=&a,\nonumber\\
F&=& 2a/3+b.
\end{eqnarray}
The $\pi N N$ and $\eta N N$
couplings are given exactly by the  formul\ae\ in the $SU(3)$ symmetry limit
with these values of $D$ and $F$.

For any given process, the matrix element of $P^{ia}$ can be expressed
as a sum of the seven parameters $a,b,d,c_1,\ldots, c_4$ times coefficients
derived from
the matrix elements of the operators. The coefficients for the axial couplings
are tabulated in Table \ref{AxiC} and those for the magnetic moments are
tabulated in Table \ref{MagC}.

\section{The Axial Couplings}\label{sec:skip}

In the large $N_c$ limit $P^{ia}$ gives the matrix
elements of the space components of the axial vector currents between baryon
states. For the octet baryons there is little ambiguity as to how to apply
this to the real world of $N_c=3$. We use the $g_A$ parameters as
conventionally defined in $\beta$-decay experiments with a normalization such
that $g_A\approx 1.26$ and $g_V=1$ for neutron decay. Experimentally, one
measures the lifetimes which are proportional to $1/(|g_V|^2+3|g_A|^2)$ and the
asymmetry which gives $g_A/g_V$. The $g_V$ parameters are taken from $SU(3)$
because of the Ademollo-Gatto theorem which states that $SU(3)$ violation in
$g_V$ occurs only in second order. When form factor
effects~\cite{Nieto},\footnote{We have used a dipole form for the axial
and vector form factors, with masses $M=1.08$~GeV for the $\Delta S=0$ axial
form factor, $M=1.25$~GeV for the $\Delta S=1$ axial form
factor~\cite{Bourquin}.  The corresponding
masses for the vector form factor are $M=0.84$~GeV and $M=0.97$~GeV.}
radiative corrections~\cite{Wilk,Garcia}, and weak magnetism are taken into
account, the values of $g_A$
obtained from the asymmetries and the rates are in general, although not
spectacularly good, agreement with each other.\footnote{In all cases, the value
of
$\left | g_A \right |$ obtained from the lifetime is greater than that obtained
from the asymmetry.}\
There is one exception: for $%
\Xi ^{-}\rightarrow\Lambda $ $\beta$-decay the two methods produce inconsistent
values of $g_A$. This can be understood from the fact that for this decay $%
g_A/g_V$ is unusually small, and the extraction of $g_A$ from the rate is
difficult both because of ambiguities in the rate and the various
corrections and because the extracted value of $g_A$ is likely to be
sensitive to the (second order) $SU(3)$ violations in $g_V$. For this decay
we use only the value of $g_A$ extracted from the asymmetry. In the case of $%
\Xi ^0$ to $\Sigma ^{+}$ decay, only the rate has been measured so we have
no choice but to take $g_A$ from the rate -- fortunately $|g_A/g_V|$ is
fairly large for this decay. For the other decays, we have combined the values
of $g_A$ from the decay asymmetry and lifetimes using scaled errors, as
recommended by the Particle Data Group~\cite{RevPP}.
The experimental values for $g_A$ are listed in
the fourth column of Table \ref{AxiD}. Apart from the $\Xi ^0\rightarrow \Sigma
^{+}$
entry, they are essentially the same as the standard values derived from the
asymmetry, except that in some cases the errors have been enlarged to
account for discrepancies between the rates and the asymmetries.

Off-diagonal elements of $P^{ia}$ connecting the decuplet baryons to the
octet baryons can be extracted from the $\pi$ decays of the decuplet. Here
there is some ambiguity in how to handle the kinematics -- in the $%
N_c\rightarrow \infty $ limit the baryon masses become infinite and static
kinematics apply, but in the real world we have to deal with finite masses.
There is no definitive answer to this problem, but Peccei \cite{Pec}
has developed a formalism that is internally consistent and compatible with
chiral symmetry. In his formalism, which we will adopt, the width of a
decuplet baryon $B^{\prime }$ decaying to a pion and an octet baryon $B$ is
\begin{equation}
\Gamma _{B^{\prime }}=\frac{g^2C(B,B^{\prime })^2(E_B+M_B){\bf q}^3}{24\pi
f_\pi
^2M_{B^{\prime }}}  \label{pece}
\end{equation}
where $E_B$ and ${\bf q}$ are the octet baryon energy and the pion
three-momentum in
the rest frame of the decaying baryon, $f_\pi $ is the pion decay constant
equal to 93 MeV, $g$ is the analog of $g_A$ for this process and $%
C(B,B^{\prime })$ is a Clebsch-Gordon coefficient $\{1,1/\sqrt{2},1/\sqrt{3}%
,1/\sqrt{2}\}$ for $\{\Delta \rightarrow N\pi ,\Sigma ^{*}\rightarrow
\Lambda\pi, \Sigma ^{*}\rightarrow \Sigma\pi, \Xi ^{*}\rightarrow \Xi\pi \}$
chosen so that all the couplings become equal in the $SU(3)$ symmetric
limit. For each decay we take the widths of the different charged states and
use the Particle Data Group~\cite{RevPP} averaging procedure to
determine $g$. These numbers are listed in the fourth column of Table
\ref{AxiD}.

\subsection{Fits to the Experimental Data}

We will perform a number of different fits to the experimental data. All the
fits have large $\chi^2$. In interpreting the results, it is important to keep
in mind that the dominant error in all the fits is {\it theoretical}; the
theoretical formul\ae\ are not as accurate as the experimental measurements.
For example, the $SU(3)$ symmetric fit to the hyperon $\beta$-decays
discussed below has $\chi^2=13.5$ for
four degrees of freedom. The large $\chi^2$ is an indication that the
experimental data show evidence for $SU(3)$ breaking in the axial vector
currents. The value of $\chi^2$ can be used to estimate the amount of $SU(3)$
breaking. If one includes, for example, a theoretical uncertainty of $\pm0.025$
(added in quadrature to the experimental errors) then $\chi^2$ is reduced to
3.6 for four degrees of freedom. This indicates that the $SU(3)$ breaking
part of the octet baryon $g_A$ (which is of order unity) is of order 0.025.
What is surprising is that $SU(3)$ breaking in the hyperon
$\beta$-decays is so small.

\subsubsection{$SU(3)$ Symmetric Fit}

We first perform a two parameter fit to the experimental data on hyperon
$\beta$-decays alone using $a$ and $b$. This is identical to a fit using only
$F$ and $D$ neglecting all $SU(3)$ breaking effects. The results are
$D=0.79\pm0.01$, $F=0.47\pm0.01$ (or equivalently $a = 0.791 \pm 0.007
$ and $b = -0.058 \pm 0.011$), and $3F-D=0.62\pm0.03$, with $\chi^2=13.5$ for
four degrees
of freedom. The results are consistent with earlier fits~\cite{Jaffe}, and the
differences are due to different treatment of the inconsistent experimental
values for $g_A$. As mentioned above, the large $\chi^2$ is an indication of
$SU(3)$ breaking. It also indicates that the nominal errors on $F$, $D$ and
$3F-D$ obtained from the fit are underestimates of the true error.

\subsubsection{$\Delta S=0$ Fit}

Before presenting the results of a full fit, it is useful to make a
preliminary investigation of the $n\rightarrow p$ and $\Sigma \rightarrow
\Lambda $ $\beta$-decays and the strong decays $\Delta \rightarrow N\pi $ and
$%
\Sigma ^{*}\rightarrow \Lambda \pi $. These $\Delta S=0$ decays depend only on
the four parameters $a$, $b$, $d$ and $c_3$ which can therefore be extracted
from this data alone. The results, $a=0.90\pm0.02$, $b=-0.24\pm0.04$,
$d=-0.05\pm0.01$ and $c_3=-0.08\pm0.01$,
are consistent with expectations. The leading parameter $a$ is of order
unity, $b\sim 1/N_c$ is small compared to $a$, $d\sim 1/N_c^2$ is quite
small and $c_3\sim \epsilon /N_c$ is consistent with a 30\% $SU(3)$ breaking
$\epsilon $ divided by $N_c=3$. The $D$ and $F$ couplings derived from this
analysis are $D=0.90\pm0.02$ and $F=0.36\pm0.02$ with
$3F-D=0.20\pm0.1$.\footnote{There are correlated errors on the
parameters in all the fits, so that the error on $F$, $D$, and $3F-D$ is
computed using the covariance matrix.}

A variant of the above is to include all six $\Delta S=0$ decays, which
can be fit using the five parameters $a$, $b$, $d$, $c_3$ and $c_4$. The
results are $a=0.894\pm0.022$, $b=-0.224\pm0.037$, $d=-0.056\pm0.010$,
$c_3=-0.079\pm0.007$,
and $c_4=0.002\pm0.014$, with $F=0.37\pm0.02$, $D=0.89\pm0.02$, and
$3F-D=0.22\pm0.09$, with $\chi^2=1.1$ for one degree of freedom.

\subsubsection{Global Fits}

Turning now to the global fit, we minimize $\chi ^2$ weighting each datum
with its experimental error. The fit has $\chi^2=7.8$ with
three degrees of freedom, and the results, labeled as Fit A, are summarized
in Tables \ref{AxiD} and \ref{AxiP}. We have tried weighting the data with
theoretical errors of various forms\footnote{For example, assuming that
there is an additional theoretical uncertainty in the formul\ae\ for the
$\Delta S=1$ decays and decuplet decays, because of the larger momentum
transfer.}\ and, although $\chi ^2$ decreases, the general character of the fit
changes very little. Note that in this fit $c_1$
and $c_4$ are very small, suggesting a fit where $c_1$ and $c_4$ are
constrained to be zero. The results of the constrained fit, labeled as Fit
B, are also listed in Tables \ref{AxiD} and \ref{AxiP}. Here $\chi ^2=9.9$
but there are now five degrees of freedom. Note that the values of $a$, $b$,
$d$, $c_2$ and $c_3$ do not change much between the constrained and
unconstrained fits.

Fit~A is shown in Fig.~\ref{fig:plot1}. The first four points are the
decuplet decays, followed by the hyperon $\beta$-decays. Rather than plot the
data points directly, we have subtracted the best $SU(3)$ symmetric fit
($a=0.791$, $b=-0.058$, $d=-0.087$) from experiment and theory. The plot shows
the deviations from $SU(3)$ symmetry of the axial couplings and the best
fit. It is clear from the plot that the $SU(3)$ breaking in the decuplet decays
is significantly larger than that in the hyperon $\beta$-decays. The hyperon
$\beta$-decays show very little $SU(3)$ breaking. Our large-$N_c$ fits
indicate that the $g_A$ values for $\Xi$-decay should be smaller than the
experimental central values in Table~\ref{AxiD}.
The Fits~A and B, as well as our $\Delta S=0$ fits all
indicate that $3F-D$ is approximately $0.27\pm0.09$, which is significantly
smaller than the $SU(3)$ symmetric fit value of $0.62\pm0.03$.

Graphically, the parameter $b$ arises from diagrams where the quark line to
which the current is attached has a spin-dependent interaction with another
quark. The parameter $c_2$ is a measure of how this interaction is
modified
when the current carrying quark line represents a strange quark. If $c_2$
were exactly equal to $-b/2$, it would mean that this spin-dependent
interaction is completely ineffective for the heavier strange quark. Since
the fits do produce a value of $c_2$ which is close to $-b/2$, we have done
another fit with three constraints, $c_1=0$, $c_4=0$ and $c_2=-b/2$. The
results of this fit, labeled as Fit C, are also shown in
Tables~\ref{AxiD} and \ref{AxiP}. By now $\chi ^2$ has increased
to 17.6 for six degrees of freedom.
Note that this doubly constrained fit has the same
four parameters that appeared in our preliminary investigation plus the
constraint $c_2=-b/2$. The parameters turn out to be almost identical with $%
a=0.87$, $b=-0.18$, $d=-0.07$, $c_3=-0.07$, $D=0.87$, $F=0.40$ and $3F-D=0.34$.
We believe that this fit is the closest to the physics. The physical
interpretation of $b$ and $c_2$ has already been discussed. To interpret $%
c_3,$ we note that in diagrams it corresponds to a non-strange quark line
carrying the current interacting, through a spin-independent gluon exchange,
with strange quarks elsewhere in the baryon. Thus, $c_3$ can be
interpreted as the effect of the strange quark mass on the average,
spin-independent color field through which the quarks propagate.

While we have achieved an apparently stable set of fits with an interesting
interpretation, we are not entirely comfortable with these results. For one
thing, we do not understand why $c_4$ and especially $c_1$ are so small.
Also, the $\Xi $ decays have been measured in only one experiment which
causes some worry --- note that in Fit A $\chi ^2$ is dominated by the
$\Xi $ decays --- and, except for the $\Delta $, there
has been no recent work on the decuplet widths. While our fit may not turn
out to be definitive, there is, nevertheless, a very clear trend. By all
indications, $3F-D\approx 0.27$ is small compared to the $SU(3)$ symmetric
value of 0.6, or the non-relativistic quark model value of one.
As a word of warning, however, we are in a parameter regime where there
are large cancellations between $3F$ and $D$ and the fractional error in $%
3F-D$ may be relatively large.

\section{Magnetic Moments}\label{sec:mag}

In the large $N_c$ limit, the baryon magnetic moments have the same
kinematic properties as the axial couplings and can be expressed in terms of
the same operators. The operator which gives the magnetic moments is $%
M^i=P^{i3}+P^{i8}/\sqrt{3}$. It is straightforward to determine the
coefficients in $M^i$. We use the measured moments of the octet baryons and
the $\Omega ^{-}$ as well as the $\Sigma ^0\Lambda $ and $\Delta N$
transition moments. The $\Delta^{++}$ magnetic moment is not included in the
fits, because of the very large experimental error.
As before, we minimize $\chi ^2$ but this time the data
are accurate enough that we have to include a theoretical error in order to
get a meaningful $\chi ^2$. The dominant error is the theoretical error in the
formul\ae\ used, not the experimental errors on the data. Guessing that the
higher order (in $SU(3)$
breaking and $1/N_c$) effects are at the few percent level we arbitrarily
add an extra error of 0.05 nuclear magnetons to each moment. The fit produces
$\chi ^2=5.3$ with four degrees of freedom but this particular value is
largely a reflection of our choice of theoretical error. The results of the
fit are listed as Fit~A in Tables \ref{MagD} and \ref{MagP}. The fit is quite
good and sizes of the various output parameters are consistent with
expectations. The plot of the deviations of the experimental data from the
$SU(3)$ symmetric values (using $a=2.87$, $b=-0.077$, and $d=-0.389$) are shown
in Fig.~\ref{fig:plot2}. The figure clearly shows that $SU(3)$ breaking in the
magnetic moments is significantly larger than in the axial current sector.
The small value of $b$ also shows that the $F/D$ ratio for the baryon magnetic
moments is very close to $2/3$.

As mentioned in the Introduction, the diagram in Fig.~\ref{fig:loop} should be
the
dominant source of $SU(3)$ violation in the magnetic moments. Using the
leading order coupling $G^{ia}$ at the meson-baryon vertices and neglecting
the baryon mass differences, this diagram can be written as
\begin{equation}
M_{\rm loop}^i=\frac 1{N_c}\epsilon ^{ijk}G^{ja}G^{kb}I^{ab}  \label{loop}
\end{equation}
where $I^{ab}$ is an antisymmetric matrix which is the result of doing the
loop integral. The Hermitian matrix $iI^{ab}$ is diagonal in a basis
corresponding to particles of definite quantum numbers. It has four zero
eigenvalues corresponding to the four neutral mesons, two equal and opposite
eigenvalues $\pm I_K$ corresponding to $K^{+}$ and $K^{-}$ and two more
equal and opposite eigenvalues $\pm I_\pi $ corresponding to $\pi ^{+}$ and $%
\pi ^{-}$. We can write
\begin{equation}
I^{ab}=\frac{I_K+2 I_\pi }3 I_0^{ab}+ \frac{I_K}{\sqrt3} I_1^{ab}+
\frac{I_\pi-I_K }{\sqrt3}J^{ab},  \label{loop1}
\end{equation}
where
\begin{eqnarray}
I_0^{ab} &=& f^{abQ},\nonumber\\
I_1^{ab} &=& f^{ab\bar Q},\\
J^{ab} &=& f^{acQ} d^{bc8} - f^{bcQ} d^{ac8} - f^{abc} d^{cQ8},\nonumber
\end{eqnarray}
and
\begin{eqnarray}
T^Q =&T^3 + \frac{1}{\sqrt3} T^8&=\left[ \begin{array}{ccc}
2/3&0&0\\
0&-1/3&0\\
0&0&-1/3 \end{array}\right], \nonumber\\
T^{\bar Q} =& T^3 - \frac{1}{\sqrt3} T^8&=
\left[ \begin{array}{ccc}
1/3&0&0\\
0&-2/3&0\\
0&0&1/3 \end{array}\right].\end{eqnarray}
$I_0^{ab}$ is an $SU(3)$ octet that transforms like the electric charge
$Q$ and gives baryon magnetic moments with $F/D=2/3$ that satisfy the
Coleman-Glashow relations~\cite{CG}. $I_1^{ab}$ is also
an $SU(3)$ octet, which transforms like the electric charge rotated by
$\pi$ in isospin space. Its contribution to the magnetic moments satisfies the
Coleman-Glashow relations, but not $F/D=2/3$. $J^{ab}$ breaks $SU(3)$ as a $
{\bf 10}\oplus \overline{\bf 10}$.
The $SU(3)$-violating part of the diagram is, therefore, a constant times
\begin{equation}
\frac 1{N_c}\epsilon ^{ijk}G^{ja}G^{kb}J^{ab}  \label{loop2}
\end{equation}
where $J^{ab}$ is completely independent of the dynamics. For example, if
the diagram is modified by putting in form factors to cut off the kaon
loops, $I_\pi-I_K $ will change but $J^{ab}$ will not. The
identities of Ref. \cite{djm2} can be used to express $\epsilon
^{ijk}G^{ja}G^{kb} J^{ab}$ in terms of the operators which occur in our
parameterization of the magnetic moments. The coefficients have the correct
scaling with $N_c$ and for the physical case of $N_c=3$, we find $c_2=-c_1/6$,
$c_3=c_1/6$ and $c_4=-c_1/6$ where $c_1$ is arbitrary and reflects the
overall scale of the diagram.
Defining deviations from the coefficients
predicted by the diagram, $\delta c_2=c_2+c_1/6$, $\delta c_3=c_3-c_1/6$ and
$\delta c_4=c_4+c_1/6$, the fit gives $c_1=-0.546\pm 0.115$, $\delta
c_2=0.011\pm 0.043$, $\delta c_3=0.004\pm 0.052$ and $\delta c_4=-0.048\pm
0.030$. The fact that $c_1$ is an order of magnitude larger than the $\delta
c_i$ is a striking indication that the symmetry breaking part of the
magnetic moments is dominated by this one diagram. Actually, the current fit
does not reflect the full content of the diagram. The expression in
Eq.~(\ref{loop2}) fails to satisfy the relation $3c_6^{\prime }=c_1^{\prime
}+c_2^{\prime }$ by a term which is explicitly of order $1/N_c$ and such
terms have been dropped in the above analysis. However, there is no harm in
keeping this extra term and doing so is equivalent to adding a term $eJ^i$
to $M^i$ where the diagram gives $e=c_1/9$ at $N_c=3$. If we add $c_1J^i/9$
to $M^i$ and redo the fit, $\chi ^2$ drops from 5.3 to 1.3 and the new
coefficients are listed as Fit~B in the tables. We take the
drop in $\chi ^2$ associated with this extra term as a further indication
that this single diagram is the dominant symmetry breaking effect in the
magnetic moments. The deviation of the baryon magnetic moments from the $SU(3)$
symmetric fit plus the loop graph of Fig.~\ref{fig:loop} is shown in
Fig.~\ref{fig:plot3}.
The residual $SU(3)$ breaking is clearly quite small.

Despite the factor of $1/N_c$ which comes from the pion couplings, $%
M_{\rm loop}^i$ is leading order in $N_c$ -- the expansion of $\epsilon
^{ijk}G^{ja}G^{kb}$ contains a term $N_cG^{ic}$. In the limit of small $m_s$%
, the symmetry breaking part of $M_{\rm loop}^i$ is also of order $\sqrt{m_s}$,
so it should not be surprising that this diagram dominates. However, when
one actually calculates the loop integrals, the numerical value of the
diagram is about a factor of two too large. The resolution of this paradox
is, no doubt, that the integrals contain virtual meson momenta up to roughly
$M_K$ which is too large for chiral perturbation theory to be valid and that
some dynamical effect is cutting off the loop integrals at a lower value of
the momenta.

\section{Conclusions}

This work has produced two clear results. The first is that the pattern of
symmetry breaking in the magnetic moments is in excellent agreement with the
pattern associated with the meson loop shown in Fig.~\ref{fig:loop}. The
magnitude of the
loop is, however, about a factor of two too big. The discrepancy is probably
due to dynamical effects that cut off the loop integral at a momentum
somewhat smaller than $M_K$. The second result is that the quantity $%
3F-D\approx 0.27$ is considerably smaller than its $SU(6)$ symmetric value of
one. This appears to be an inescapable result of our fits and is, of course,
important in the analysis of spin-dependent deep inelastic scattering.
For
the axial couplings we also found $c_2\approx -b/2$ which suggests that the
strange quark is heavy enough that its spin-dependent interactions are
rather strongly suppressed. Beyond that, however, the fit to the axial
couplings, while good and seemingly stable, is rather unsatisfying from a
theoretical point of view. In particular, we do not know why $c_4$ and
especially $c_1$ are so small. Further progress on the axial couplings
probably awaits better data, which is likely to be years away, or a better
understanding of large $N_c$ baryon dynamics, which might produce a simple
reason for the smallness of $c_1$.

\acknowledgements

This work was supported in part by the Department of Energy under Grant No.
DOE-FG03-90ER40546. E.J. was supported in part by the NYI program, through
Grant No.~PHY-9457911 from the National Science Foundation, and by the Alfred
P.~Sloan Foundation. A.M. was supported in part by the PYI program, through
Grant No.~PHY-8958081 from the National Science Foundation.

\appendix

\section{Higher Order Operators}

The point of this appendix is to show that the truncation scheme is
consistent in the sense that matrix elements of the operators that have been
dropped are always smaller, for large $N_c$, than the corresponding matrix
elements of at least one of the operators that has been retained. This has
already been done for the singlet and octet operators -- the corresponding
analysis for the ${\bf 10}\oplus \overline{\bf 10}$ and ${\bf 27}$ follows.
Up to terms which
simply renormalize the symmetrical couplings and irrelevant constants, the
operators that we keep are
\begin{eqnarray}
O_1&=&\left\{G^{ia},\frac{N_s}{N_c}\right\}, \nonumber\\
O_2&=&\left\{T^a,\frac{J_s^i}{N_c}\right\},
\end{eqnarray}
where the scaling with $N_c$ has been made explicit, and we drop the
operators
\begin{eqnarray}
O_3 &=&\frac{J^i}{N_c}\left\{T^a,\frac{N_s}{N_c}\right\}, \nonumber \\
O_4 &=&\left\{\frac{J^kG^{ka}}{N_c},\frac{J_s^i}{N_c}\right\},  \\
O_5 &=&\left\{\frac{G^{ia}}{N_c},\frac{J^kJ_s^k}{N_c}\right\}.  \nonumber
\end{eqnarray}
First consider the operators $O_1$, $O_4$ and $O_5$, all of which contain the
basic operator $G^{ia}$. Since $J_s^2=(N_s/2)(N_s/2+1)$, $J_s^i$ cannot have
matrix elements larger than $[(N_s/2)(N_s/2+1)]^{1/2}$ and it is clear that
for low spin states any matrix element of $O_4$ or $O_5$ is of order $%
N_c^{-1}$ times the corresponding matrix element of $O_1$. Dropping $O_4$
and $O_5$ is therefore justified by the fact that these operators are always
small compared to $O_1$. Now consider the matrix elements of $O_2$ and $O_3$%
, both of which contain the basic operator $T^a$. The comparison of these
operators is simplified by noting that $O_3$ has matrix elements only
between states of the same spin and that the matrix element of $J_s^i$
between states of equal spin is equal to the matrix element of $%
J^iJ^kJ_s^k/J^2$. Therefore the statement that $O_3$ is small compared to $%
O_2$ is equivalent to the statement that, for low spin states, $%
\{T^a,N_s/N_c\}$ is small compared to $\{T^a,J_s^kJ^k/J^2\}$. We will
explicitly carry out the calculation for the experimentally interesting case
of spin-$1/2$ -- other cases can be studied in a similar way. The
structure of the baryon multiplet is such that: (i) the isospin of a baryon
is equal to the total angular momentum (spin) of the up and down quarks and
(ii) the total angular momentum (spin) of the strange quarks is $N_s/2$. It
follows that: (i) there are no spin-$1/2$ baryons with $N_s>(N_c+1)/2$
and (ii) a spin-$1/2$ baryon must have isospin $(N_s-1)/2$ or $(N_s+1)/2$
where either possibility is allowed for $1\leq N_s\leq (N_c-1)/2$ and only the
smaller isospin is allowed when $N_s=(N_c+1)/2$.  The case $N_s=0$ also is
exceptional, with only the larger isospin allowed, $I= 1/2$.
A straightforward
calculation then shows that $J^kJ_s^k/J^2$ is equal to $-N_s/3$ for the
higher isospin and $(N_s+2)/3$ for the lower isospin. Note that, unlike the
matrix elements of $N_s$, the matrix elements of $J_s^kJ^k/J^2$ can have
different signs giving rise to the possibility of cancellations between the
two terms in $\{T^a,J_s^kJ^k/J^2\}$. If the $SU(3)$ index $a$ is $1,2,3$ or $%
8$, then neither $O_2$ nor $O_3$ can change the isospin of a state, so there
can be no cancellation for any $N_s$ and the matrix elements of $O_3$
are down by $N_c^{-1}$ relative to the matrix elements of $O_2.$ However, in
the case of strangeness changing operators, a cancellation can occur. There
are two types of strangeness changing matrix elements, those with $\Delta
N_s=2\Delta I$ where the state with the largest number of strange quarks
also has the larger isospin and those with $\Delta N_s=-2\Delta I$ where the
state with the largest number of strange quarks has the smaller isospin. For
the $\Delta N_s=2\Delta I$ transitions, the two terms in $O_2$ have the same
sign and, independent of the value of $N_s$, the matrix element of $O_3$ is
always smaller than that of $O_2$ by a factor of $1/N_c$. However, for the $%
\Delta N_s=-2\Delta I$ transitions there is a cancellation and, for large $%
N_s$, the matrix element of $O_3$ is smaller than that of $O_2$ by a
factor of order $N_s/N_c$, i.e. a factor which is of order unity when $%
N_s\sim N_c$. However, for these transitions the matrix elements of $O_1$
are large compared to the matrix elements of either $O_2$ or $O_3$. To see
this, we note that the matrix element of $G^{ia}$ between states of the same
spin is equal to the matrix element of $J^iJ^kG^{ka}/J^2$ and use the
identity
\begin{equation}
J^kG^{ka}=\frac 12d^{abc}T^bT^c+\frac 14\left(\frac{N_c}3+1\right)T^a
\end{equation}
which is a linear combination of two of the (0,adj) $SU(6)$ identities
listed in Ref.\cite{djm2}. Taking matrix elements of this identity and
working out the Clebsch-Gordon coefficients yields
\begin{equation}
J^kG^{ka}/J^2=\frac 23\left(T^a+\frac 12\{N_s,T^a\}\right)
\end{equation}
for $\Delta N_s=-2\Delta I$ transitions, where this relation is understood
to be valid only for matrix elements of strangeness changing operators between
spin-$1/2$ baryons. It follows immediately that for $N_s\sim N_c$ the
strangeness changing matrix elements of $O_3$ are order $N_c^{-2}$ times the
corresponding matrix elements of $O_1$. The conclusion of these rather
lengthy calculations is that in all cases matrix elements of $O_{3\text{, }%
}O_4$ and $O_5$ are smaller than a corresponding matrix element of $O_1$ or $%
O_2$ by at least a factor of $N_c^{-1}$.

\newpage\onecolumn

\begin{table}[tbp] \centering
\caption{Coefficients for axial couplings. \label{AxiC}}
\begin{tabular}{cccccccc}
& $a$ & $b$ & $d$ & $c_1$ & $c_2$ & $c_3$ & $c_4$ \\ \hline
$\Delta N$ & $-2$ & $0$ & $9/2$ & $0$ & $0$ & $0$ & $0$ \\
\noalign{\smallskip}
$\Sigma ^{*}\Lambda $ & $-2$ & $0$ & $9/2$ & $0$ & $0$ & $-4$ & $0$ \\
\noalign{\smallskip}
$\Sigma ^{*}\Sigma $ & $-2$ & $0$ & $9/2$ & $0$ & $0$ & $-4$ & $8$ \\
\noalign{\smallskip}
$\Xi ^{*}\Xi $ & $-2$ & $0$ & $9/2$ & $0$ & $0$ & $-8$ & $4$ \\
\noalign{\smallskip}
$np$ & $5/3$ & $1$ & $0$ & $0$ & $0$ & $0$ & $0$ \\
\noalign{\smallskip}
$\Sigma \Lambda $ & $\sqrt{2/3}$ & $0$ & $0$ & $0$ & $0$ & $\sqrt{8/3}$ & $%
0$ \\
\noalign{\smallskip}
$\Lambda p$ & $-\sqrt{3/2}$ & $-\sqrt{3/2}$ & $0$ & $-\sqrt{3/2}$ & $%
-\sqrt{3/2}$ & $-\sqrt{3/2}$ & $-\sqrt{3/2}$ \\
\noalign{\smallskip}
$\Sigma n$ & $1/3$ & $-1$ & $0$ & $1/3$ & $-1$ & $1/3$ & $1/3$ \\
\noalign{\smallskip}
$\Xi \Lambda $ & $1/\sqrt{6}$ & $\sqrt{3/2}$ & $0$ & $1/\sqrt{6}$ & $%
\sqrt{3/2}$ & $\sqrt{3/2}$ & $7/\sqrt6$ \\
\noalign{\smallskip}
$\Xi \Sigma $ & $5/\sqrt{18}$ & $1/\sqrt{2}$ & $0$ & $5/\sqrt{18}$ &
$1/\sqrt{2}$
& $5/\sqrt2$ & $1/\sqrt{2}$%
\end{tabular}
\end{table}

\begin{table}[tbp] \centering
\caption{Coefficients for magnetic moments. \label{MagC}}
\begin{tabular}[t]{cccccccc}
& $a$ & $b$ & $d$ & $c_1$ & $c_2$ & $c_3$ & $c_4$ \\ \hline
$p$ & $1$ & $1$ & $0$ & $0$ & $0$ & $0$ & $0$ \\
$n$ & $-2/3$ & $0$ & $0$ & $0$ & $0$ & $0$ & $0$ \\
$\Lambda $ & $-1/3$ & $0$ & $0$ & $-2/3$ & $-2/3$ & $-2/3$ & $0$ \\
$\Sigma ^{+}$ & $1$ & $1$ & $0$ & $2/9$ & $-2/3$ & $2$ & $-2/3$ \\
$\Sigma ^{-}$ & $-1/3$ & $-1$ & $0$ & $2/9$ & $-2/3$ & $-2/3$ & $2/3$ \\
$\Sigma ^0\Lambda $ & $-1/\sqrt3$ & $0$ & $0$ & $0$ & $0$ & $-2/\sqrt3$ & $%
0$ \\
$\Xi ^0$ & $-2/3$ & $0$ & $0$ & $-8/9$ & $-4/3$ & $-8/3$ & $0$ \\
$\Xi ^{-}$ & $-1/3$ & $-1$ & $0$ & $-8/9$ & $-4/3$ & $-4/3$ & $-8/3$ \\
$p\Delta ^{+}$ & $\sqrt8/3$ & $0$ & $-3/\sqrt2$ & $0$ & $0$ & $0$ & $0$ \\
$\Omega ^{-}$ & $-1$ & $-3$ & $0$ & $-2$ & $-6$ & $-6$ & $-6$ \\
$\Delta ^{++}$ & $2$ & $6$ & $0$ & $0$ & $0$ & $0$ & $0$%
\end{tabular}
\end{table}

\begin{table}[tbp] \centering
\caption{Axial coupling fits. \label{AxiD}}
\begin{tabular}{ccccccc}
Decay & Experiment & Coupling & Average & Fit A & Fit B &Fit C\\ \hline
$\Delta\rightarrow N\pi$&& & -2.04 $\pm $ 0.01 & -2.04 & -2.04 & -2.03 \\
$\Sigma ^{*}\rightarrow\Lambda\pi $&& & -1.71 $\pm $ 0.03 & -1.75 & -1.73 &
-1.74 \\
$\Sigma ^{*}\rightarrow\Sigma\pi $ && & -1.60 $\pm $ 0.13 & -1.62 & -1.73 &
-1.74 \\
$\Xi ^{*}\rightarrow\Xi\pi $&& & -1.42 $\pm $ 0.04 & -1.40 & -1.42 & -1.45 \\
\noalign{\medskip}
$n\rightarrow p\ell\nu$&$e$ &1.2711 $\pm $ 0.002 &1.2664 $\pm$ 0.0065&
1.266 & 1.266 & 1.271\\
&$g_A/g_V$& 1.2573 $\pm $ 0.0028&&&\\
\noalign{\medskip}
$\Sigma\rightarrow\Lambda\ell\nu$&$\Sigma^+\rightarrow e$ & 0.601 $\pm $ 0.015
&0.602 $\pm $ 0.014& 0.602 & 0.596 & 0.590\\
&$\Sigma^-\rightarrow e$ & 0.624 $\pm $ 0.079 &&& \\
\noalign{\medskip}
$\Lambda\rightarrow p\ell\nu$&$e$ & -0.906 $\pm $ 0.024 &-0.890 $\pm$ 0.015&
-0.896 & -0.901 & -0.867\\
&$g_A/g_V$ & -0.879 $\pm $ 0.018 &&& \\
&$\mu$ & -0.977 $\pm $ 0.180 &&& \\
\noalign{\medskip}
$\Sigma\rightarrow n\ell\nu$&$e$ & 0.348 $\pm $ 0.030 &0.341 $\pm$ 0.015 &
0.339 & 0.342 & 0.352\\
&$g_A/g_V$& 0.340 $\pm $ 0.017 &&& \\
&$\mu$& 0.309 $\pm $ 0.071 &&& \\
\noalign{\medskip}
$\Xi\rightarrow \Lambda\ell\nu $&$e$ & 0.428 $\pm $ 0.049 &0.306 $\pm$
0.061 & 0.220 & 0.178 & 0.158\\
&$g_A/g_V$ & 0.306 $\pm $ 0.061 &&& \\
&$\mu$ & 1.010 $\pm $ 0.776 &&& \\
\noalign{\medskip}
$\Xi\rightarrow \Sigma\ell\nu $& $e$ & 0.929 $\pm $ 0.112 & 0.929 $\pm $
0.112 & 0.718 & 0.718 & 0.703
\end{tabular}
\end{table}

\begin{table}[tbp] \centering
\caption{Best fit parameters for axial couplings. \label{AxiP}}
\begin{tabular}{cccc}
& Fit A & Fit B & Fit C \\ \hline
$a$   & 0.882 $\pm $ 0.021 &  0.885  $\pm $ 0.013 &  0.868 $\pm$ 0.011 \\
$b$   &-0.203 $\pm $ 0.036 & -0.209  $\pm $ 0.022 & -0.175 $\pm$ 0.018 \\
$d$   &-0.061 $\pm $ 0.009 & -0.059  $\pm $ 0.006 & -0.066 $\pm$ 0.005 \\
$c_1$ &-0.022 $\pm $ 0.024 &  0                   &  0                 \\
$c_2$ & 0.132 $\pm $ 0.038 &  0.136  $\pm $ 0.020 &        $-b/2$      \\
$c_3$ &-0.072 $\pm $ 0.006 & -0.077  $\pm $ 0.004 & -0.073 $\pm$ 0.004 \\
$c_4$ & 0.016 $\pm $ 0.012 &  0                   &  0                 \\
$F$   & 0.39  $\pm $ 0.02  &  0.38   $\pm $ 0.014 &  0.40  $\pm$ 0.01  \\
$D$   & 0.88  $\pm $ 0.02  &  0.89   $\pm $ 0.013 &  0.87  $\pm$  0.01 \\
$3F-D$& 0.27  $\pm $ 0.09  &  0.26   $\pm $ 0.05  &  0.34  $\pm$  0.05
\end{tabular}
\end{table}

\begin{table}[tbp] \centering
\caption{Magnetic moment fits. \label{MagD}}
\begin{tabular}[t]{cccc}
& Experiment & Fit A & Fit B \\ \hline
$p$                 &  2.793 $\pm $ 0.000 &  2.842 &  2.801 \\
$n$                 & -1.913 $\pm $ 0.000 & -1.871 & -1.915 \\
$\Lambda $          & -0.613 $\pm $ 0.004 & -0.581 & -0.595 \\
$\Sigma ^{+}$       &  2.458 $\pm $ 0.010 &  2.449 &  2.462 \\
$\Sigma ^{-}$       & -1.160 $\pm $ 0.025 & -1.074 & -1.131 \\
$\Sigma ^0\Lambda $ & -1.610 $\pm $ 0.080 & -1.520 & -1.511 \\
$\Xi ^0$            & -1.250 $\pm $ 0.014 & -1.288 & -1.261 \\
$\Xi ^{-}$          & -0.651 $\pm $ 0.003 & -0.619 & -0.635 \\
$p\Delta ^{+}$      &  3.230 $\pm $ 0.100 &  3.530 &  3.530 \\
$\Omega ^{-}$       & -1.940 $\pm $ 0.220 & -2.166 & -2.094 \\
$\Delta ^{++}$      &  5.6   $\pm $ 1.9   &  5.822 &  5.861
\end{tabular}
\end{table}

\begin{table}[tbp] \centering
\caption{Best fit parameters for magnetic moments. \label{MagP}}
\begin{tabular}{ccc}
& Fit A & Fit B \\ \hline
$a$          & 2.807 $\pm $ 0.061 & 2.782 $\pm $ 0.058 \\
$b$          & 0.036 $\pm $ 0.059 & 0.080 $\pm $ 0.053 \\
$d$          &-0.417 $\pm $ 0.071 &-0.428 $\pm $ 0.071 \\
$c_1$        &-0.546 $\pm $ 0.115 &-0.545 $\pm $ 0.106 \\
$c_2$        & 0.102 $\pm $ 0.046 & 0.037 $\pm $ 0.043 \\
$\delta c_2$ & 0.011 $\pm $ 0.043 &-0.053 $\pm $ 0.045 \\
$c_3$        &-0.087 $\pm $ 0.037 &-0.083 $\pm $ 0.036 \\
$\delta c_3$ & 0.004 $\pm $ 0.052 & 0.008 $\pm $ 0.050 \\
$c_4$        & 0.043 $\pm $ 0.038 & 0.042 $\pm $ 0.036 \\
$\delta c_4$ &-0.048 $\pm $ 0.030 &-0.049 $\pm $ 0.030 \\
\end{tabular}
\end{table}

\vfill\break\eject

\epsffile{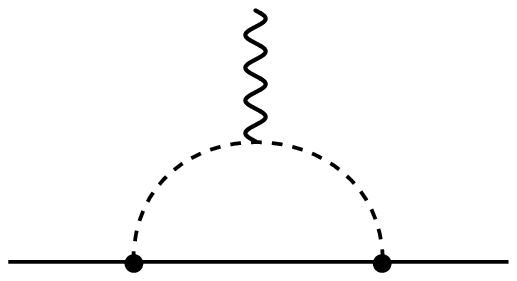}
\begin{figure}
\caption{Leading chiral correction to the baryon magnetic moments. The solid
line is the baryon, and the dashen line is the meson. The graph is
of order $M_K \sim \protect\sqrt{m_s}$.}
\label{fig:loop}
\end{figure}

\moveright1in\hbox{
\epsfxsize=4in
\epsffile{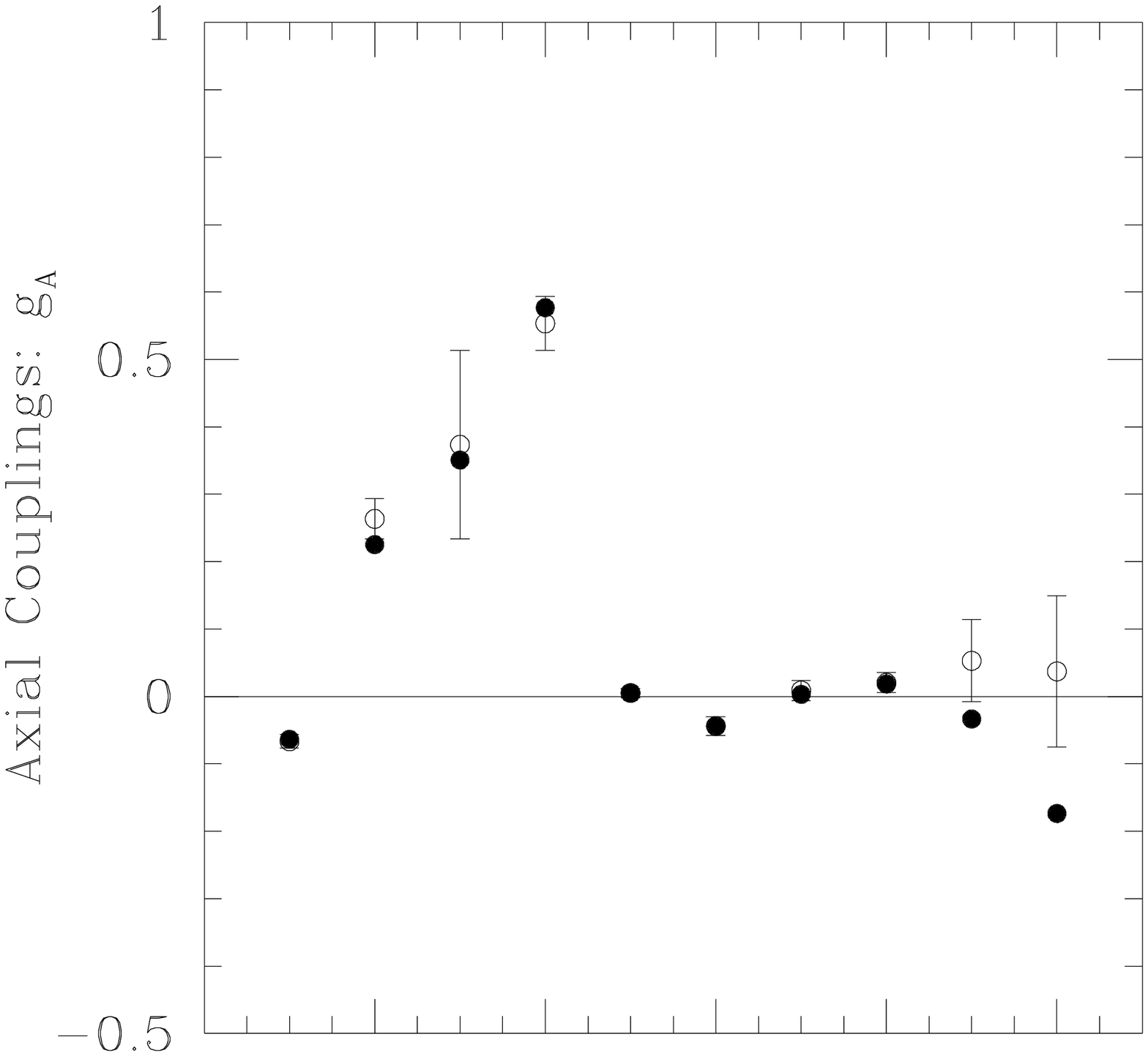}
}
\begin{figure}
\caption{Deviation of the axial couplings from the best $SU(3)$-symmetric
fit. The open circles are the experimental data, and the filled circles are the
values from Fit~A discussed in the text. The points plotted are (from left to
right) $\Delta\rightarrow N$, $\Sigma^* \rightarrow \Lambda$, $\Sigma^*
\rightarrow \Sigma$, $\Xi^* \rightarrow \Xi$, $n\rightarrow p$, $\Sigma
\rightarrow \Lambda$, $\Lambda \rightarrow p$, $\Sigma \rightarrow n$, $\Xi
\rightarrow \Lambda$, and $\Xi \rightarrow \Sigma$.}
\label{fig:plot1}
\end{figure}

\moveright1in\hbox{
\epsfxsize=4in
\epsffile{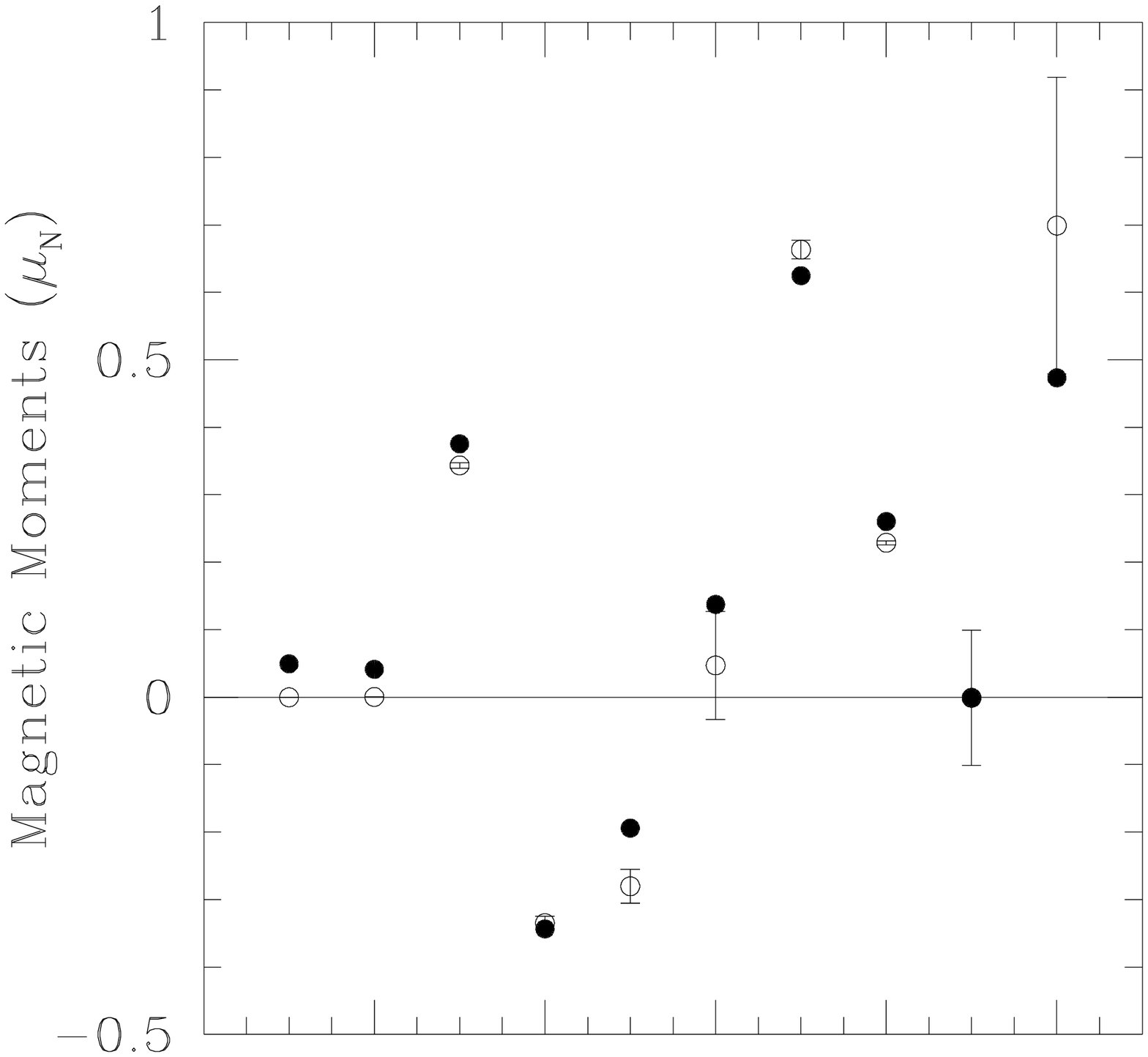}
}
\begin{figure}
\caption{Deviation of the magnetic moments from the best $SU(3)$-symmetric
fit. The open circles are the experimental data, and the filled circles are the
values from Fit~A discussed in the text. The order of the magnetic moments is
the same as in Table~\protect\ref{MagD}. The $\Delta^{++}$ magnetic moment
has not been plotted, since the experimental value has a very large error.}
\label{fig:plot2}
\end{figure}

\moveright1in\hbox{
\epsfxsize=4in
\epsffile{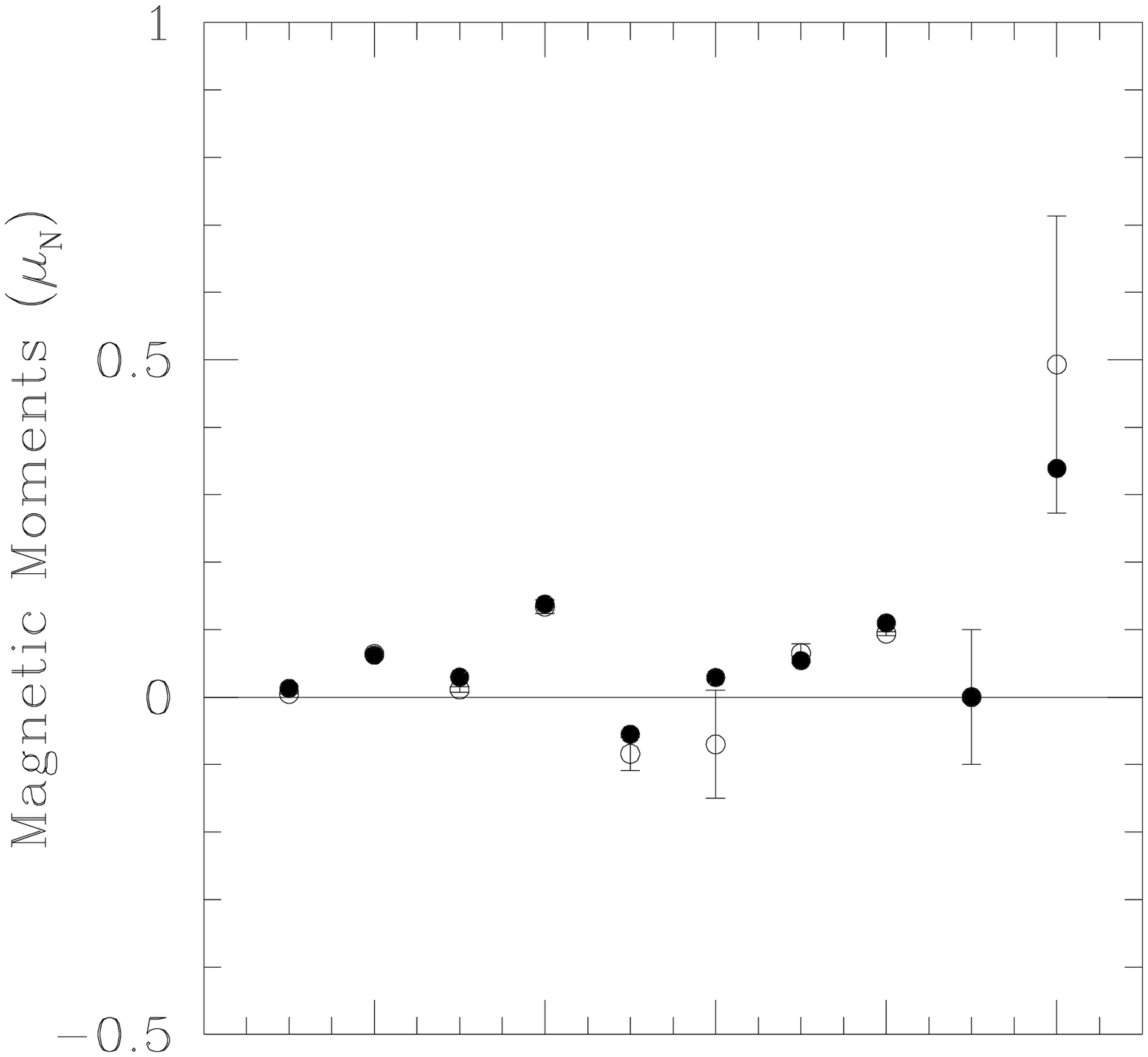}
}
\begin{figure}
\caption{Deviation of the magnetic moments from the best $SU(3)$-symmetric
fit plus the chiral loop diagram of Fig.~\protect\ref{fig:loop}. The deviations
should be compared with those in Fig.~\protect\ref{fig:plot2}.}
\label{fig:plot3}
\end{figure}

\end{document}